\documentclass[reprint,aps,prd,superscriptaddress,preprintnumbers,nofootinbib]{revtex4-1}

\usepackage{mathrsfs} 
\usepackage{amsfonts}
\usepackage{amsmath}
\usepackage{amsthm}
\usepackage{amssymb}
\usepackage{array}
\usepackage{dcolumn}
\usepackage[normalem]{ulem}
\usepackage[svgnames]{xcolor}
\usepackage[hidelinks]{hyperref}
\hypersetup{
    colorlinks=true,
    linkcolor=NavyBlue,
    filecolor=NavyBlue,
    urlcolor=NavyBlue,
    citecolor=NavyBlue,
    }

\usepackage{graphics}
\usepackage{graphicx}
\usepackage[caption=false]{subfig}
\usepackage{adjustbox}
\usepackage{breqn}
\usepackage{fontawesome} 
\usepackage{slashed}

\newcommand\lsim{\lesssim}
\newcommand\gsim{\gtrsim}

\def\be{\begin{equation}}
\def\ee{\end{equation}}
\def\bea{\begin{eqnarray}}
\def\eea{\end{eqnarray}}

\newcount\Comments  
\newcommand{\kibitz}[2]{\ifnum\Comments=1\textcolor{#1}{#2}\fi}

\makeatletter
\let\cat@comma@active\@empty
\makeatother

\begin{document}

\hfill \preprint{LA-UR-24-20773}

\title{Detecting Boosted Dark Photons with Gaseous Detectors}

\author{Michael L. Graesser}
\affiliation{Theoretical Division, Los Alamos National Laboratory, Los Alamos, NM 87545, USA}

\author{R. Andrew Gustafson}
\affiliation{Center for Neutrino Physics, Department of Physics, Virginia Tech University, Blacksburg, Virginia 24601, USA}

\author{Kate Hildebrandt}
\affiliation{School of Physics and Astronomy, University of Minnesota, Minneapolis, MN 55455, USA}

\author{Varun Mathur}
\affiliation{Center for Neutrino Physics, Department of Physics, Virginia Tech University, Blacksburg, Virginia 24601, USA}

\author{Ian M.~Shoemaker}
\affiliation{Center for Neutrino Physics, Department of Physics, Virginia Tech University, Blacksburg, Virginia 24601, USA}


\begin{abstract}
{\centering{\href{https://github.com/AndrewGustafson/BoostedDarkPhoton}{\large\color{BlueViolet}\faGithub}}  \\}
We search for indirect signals of $\mathscr{O}$(keV) dark matter annihilating or decaying into $\mathscr{O}$(eV) dark photons. These dark photons will be highly boosted, predominantly transversely polarized, have decay lengths larger than the Milky Way, and can be absorbed by neutrino or dark matter experiments at a rate dependent on the photon-dark photon kinetic mixing parameter and the optical properties of the experiment. We show that current experiments can not probe new parameter space, but future large-scale gaseous detectors with low backgrounds (i.e. CYGNUS, NEXT, PANDAX-III) may be sensitive to this signal when the annihilation cross section is especially large.
\end{abstract}

\maketitle

\section{Introduction} \label{sec:introduction}
Dark photons are a renormalizable extension of the Standard Model (SM), with a Lagrangian
\begin{equation}
    \mathscr{L} \supset -\frac{1}{4} F^{\prime}_{\mu \nu} F^{\prime \mu \nu} +\frac{1}{2} \epsilon F^{\prime}_{\mu \nu}F^{\mu \nu} +\frac{1}{2} m_{A^\prime}^2 A^{\prime \mu}A^\prime_{\mu} \label{eq:DP_Lagrangian}~.
\end{equation}
Here $A^\prime$ and $F^\prime$ are the dark field and dark field strength respectively, $F$ is the Standard Model (SM) photon field strength, $m_{A^\prime}$ is the mass of the dark photon, and $\epsilon$ is the kinetic mixing (first explored in \cite{Holdom:1985ag}, see \cite{PhysRevD.104.095029} for a recent review). This kinetic mixing gives the dark photon a coupling to SM electromagnetic currents, leading to observable interactions. The dark photon mass is assumed throughout to be of a St\"uckelberg-type.

In this paper we focus on models in which cold dark matter (CDM) is partially comprised of fermionic $\chi$ particles that are charged particles of the dark sector with Lagrangian
\begin{equation}
    \mathscr{L} \supset \Bar{\chi} (i \partial_{\mu} \gamma^{\mu} - m_{\chi}) \chi - g_{D} A^{\prime}_{\mu} \Bar{\chi} \gamma^{\mu} \chi, \label{eq:chi_Lagrangian}
\end{equation}
where $m_{\chi}$ is the mass of the fermion, and $g_{D}$ is the dark electric charge of $\chi$.

We consider a scenario where both $\chi$ and its anti-particle $\Bar{\chi}$ are present in the Milky Way. If $m_\chi \gg m_{A}$, then the annihilation of a $\chi$ and $\Bar{\chi}$ will produce ultra-relativistic dark photons, allowing them to be seen in terrestrial, low-background experiments (i.e. neutrino experiments). Similar analyses have been done using other decay/annihilation products of dark matter \cite{Huang:2013xfa,Agashe:2014yua,Cherry:2015oca} or with other astrophysical sources \cite{Agashe:2020luo}.

This scenario allows us to explore the signal from low-mass dark matter (DM) that couple to dark photons. Until recent years, this low-mass DM was relatively unconstrained by direct detection experiments.  The difficulty low-mass DM presents is that the recoil energy deposited is proportional to the DM mass, typically falling below the detector threshold for masses less than a few GeV. While low-threshold detector technologies have made advances in recent years, new strategies and materials have great promise to lead the field in constraining low-mass DM~\cite{Essig:2011nj, Essig:2012yx, Graham:2012su, An:2014twa, Essig:2015cda, Hochberg:2015pha, Derenzo:2016fse, Bloch:2016sjj, Hochberg:2016ntt, Hochberg:2016ajh, Kouvaris:2016afs, Essig:2017kqs, Budnik:2017sbu, Bunting:2017net, Knapen:2017ekk, Hochberg:2017wce, Hertel:2018aal, Dolan:2017xbu, Bringmann:2018cvk, Emken:2019tni, Essig:2019kfe, Ema:2018bih, Bell:2019egg, Trickle:2019ovy,Trickle:2019nya, Griffin:2019mvc, Baxter:2019pnz, Kurinsky:2019pgb, Catena:2019gfa, Griffin:2020lgd, Flambaum:2020xxo, Bell:2021zkr}.  

The layout of this paper is as follows: In Sec. II, we will discuss the distribution of $\chi$ within the Milky Way in light of annihilations and the corresponding dark photon flux. In Sec. III we describe the interaction of dark photons with matter, in particular how the optical properties of experiments can enhance or suppress dark photon absorption. In Sec. IV we show results from existing and projected experiments. Sec. V covers existing constraints on this model, while Sec. VI discusses a similar signal arising from decaying dark matter.

\section{$\chi$ Distribution and Dark Photon Flux} \label{sec:DM Density}
In this work, we consider the distributions of $\chi$ and $\Bar{\chi}$ to be identical, and will define $\rho_{\chi}$ to be the combined mass density distribution of $\chi$ and $\Bar{\chi}$ (i.e. the density distribution of just $\chi$ is $\rho_{\chi}/2$). In this paper, we consider large $\chi$ annihilation cross sections, so this necessarily makes our distribution time-dependent. 
Annihilations deplete the energy distribution of $\chi$ according to \cite{Kaplinghat:2000vt,Chan:2016mhw}
\begin{equation}
\frac{\partial \rho_\chi}{\partial t} = - \frac{\langle \sigma v \rangle}{m_\chi}\rho^2_\chi \label{eq:AnnTimeEvo}
\end{equation}
for $s-$wave annihilation.
We take $\rho_{\chi}$ to be proportional to the NFW profile \cite{1996ApJ...462..563N} at every location in space, with a proportionality constant which depends on time (see Appendix \ref{appendix: DM Time Evolution} for more comments on this assumption),

\begin{equation}
    \rho_{\chi}(r,t) = f_{\chi}(t) \rho_{\mathrm{NFW}}(r) \label{eq:rho_chi_decomp}
\end{equation}

\begin{equation}
    \rho_{\mathrm{NFW}}(r) = \rho_{s} \frac{r_s}{r}  \bigg( 1 + \frac{r}{r_s} \bigg)^{-2} \label{eq:NFW_dens}
\end{equation}
where $\rho_{s} =$ 0.184 GeV $\mathrm{cm}^{-3}$ and $r_{s} =$ 24.42 kpc \cite{cirelli2011pppc}. If we let $f_{\chi}(t = 0) = f_{i}$, then the time evolution of this proportionality constant is 
\begin{equation}
    f_{\chi}(t) = \bigg( \frac{1}{f_{i}} + \frac{\langle \sigma v \rangle t}{m_{\chi}} \frac{\int \rho_{\mathrm{NFW}}^2(\mathbf{x}) dV}{\int \rho_{\mathrm{NFW}}(\mathbf{x}) dV} \bigg)^{-1}, \label{eq:chi_frac}
\end{equation}
Where $\langle \sigma v \rangle$ is the present-day thermally averaged cross section and the volume integral is taken over a sphere 60 kpc in radius matching the upper limit from \cite{cirelli2011pppc} \footnote{Extending past this radius will only have a logarithmic increase on the total dark matter mass.}. We can use this to find the present-day ($t = t_{0}$) density of $\chi$, and compute the mono-energetic flux of dark photons using a line-of-sight integral \cite{cirelli2011pppc}

\begin{equation}
    \Phi_{A'} = \frac{1}{4\pi} \frac{\langle \sigma v \rangle}{2 m^2_{\chi}} \int d\Omega \int_{\mathrm{LOS}} dx \rho^2_{\chi}(x,\theta,t_0), \label{eq:Flux_Eq}
\end{equation}
where $x$ is the distance from the Sun, and $\theta$ is the angle between our line-of-sight and a line pointed towards the galactic center. We can relate this to $r$ by $r = \sqrt{x^2 + r_{\odot}^2 - 2 x r_{\odot} \cos(\theta)}$, with $r_{\odot}$ being the galactic radius of the sun. Technically, the time used for the density $t = t_{0} - \frac{x}{v}$ to account for the propagation time of the dark photons, but this effect is small compared to cosmological time scales necessary for appreciable annihilation. We can decompose this flux using Eq. \ref{eq:rho_chi_decomp}, and then find the annihilation cross section which provides the largest flux, namely

\begin{equation}
    \langle \sigma v \rangle_{\max} = \frac{m_{\chi}}{f_{i} t_{0}} \frac{\int \rho_{\mathrm{NFW}}(\mathbf{x}) dV}{\int \rho^2_{\mathrm{NFW}}(\mathbf{x}) dV}. \label{eq:sigmav_max}
\end{equation}

\begin{figure}[hbt]
    \includegraphics[width=1\columnwidth]{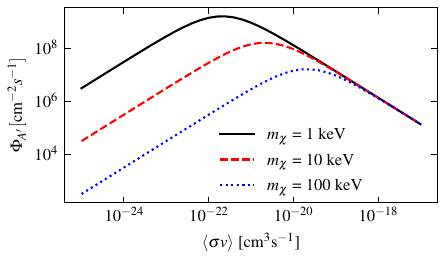}
    \caption{Flux of dark photons at Earth vs. $\langle \sigma v \rangle$ obtained by computing Eq. \ref{eq:Flux_Eq} for different $\chi$ masses, all with $f_{i}$ = 0.1. We can clearly see the optimum $\langle \sigma v \rangle$ for each mass.
    \label{fig:Flux_vs_sigmav}}
\end{figure}

This result is seen clearly in Fig. \ref{fig:Flux_vs_sigmav}. For much of the paper, we will assume this optimum annihilation cross section in order to make strong statements about which experiments are unable to probe this signal. However, if we were to take other values of $\langle \sigma v \rangle$, the flux roughly scales as

\begin{equation}
    \log(\Phi_{A'}) \sim -\bigg|\log \bigg(\frac{\langle \sigma v \rangle} {\langle \sigma v \rangle_{\max}} \bigg) \bigg|.
\end{equation}

In this work 
we remain agnostic about the origin of $\chi$ in the early Universe. 
While thermal relics are already excluded for most dark matter coupled to dark photons \cite{delaVega:2023dmw} and for warm dark matter below 5 keV \cite{Dekker:2021scf}, a small window 
remains. We note that the scenario considered here lies outside of this region, as the values of $\langle \sigma v \rangle_{\max}$ are simply too large.

In principle, the dark photon flux at our detector would also be shaped by dark photon decays en route to Earth and attenuation when passing through Earth. However, for the parameters considered in this paper, the decay length is much larger than galactic scales and the cross section is not large enough for significant attenuation to occur.

\section{Dark Photon Interactions with Matter} \label{sec:DP Interaction}
The absorption rate of dark photons is affected by the optical properties of the detector. The in-medium photon propagator leads to an effective mixing parameter of

\begin{equation}
    \epsilon_{\mathrm{eff}} = \frac{m^2_{A'}}{m^{2}_{A'} - \Pi_{T/L}} \times \epsilon, \label{eq:eps_eff}
\end{equation}
where $\Pi_{T}$ ($\Pi_{L}$) is the transverse (longitudinal) polarization tensor of the medium \cite{An:2013yfc,An:2013yua,PhysRevD.26.1394}. In our work, we are interested in transversely polarized dark photons, as in the annihilation of $\chi$ particles at the Galactic Center these are the dominant annihilation products in the limit $E_\chi \simeq m_\chi \gg m_{A'}$. This result is demonstrated in Appendix \ref{app:dark-photon-polarizations}. For an isotropic and non-magnetic (the relative permeability is 1) material, we can relate the polarization tensor to the (complex) index of refraction $n_{\mathrm{ref}}$ via \cite{PhysRevD.26.1394}~(see also Appendix A of \cite{Hochberg:2015fth} or \cite{ Coskuner:2019odd})
\begin{equation}
    \Pi_{T}(\omega) = \omega^2 (1 - n_{\mathrm{ref}}^2).
\end{equation}

\begin{figure}[hbt]
    \includegraphics[width=1\columnwidth]{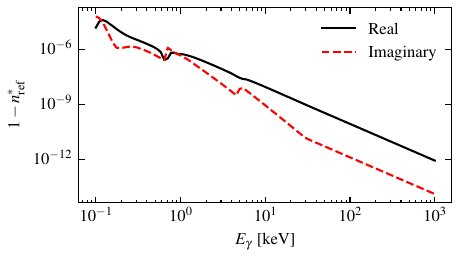}
    \caption{The real and imaginary part of $1-n^{*}_{\mathrm{ref}}$ for a xenon gas at $\rho_{Xe} = 5\, \mathrm{kg} \, \mathrm{m}^{-3}$. Note that for this plot, we are using the complex conjugate of the index of refraction, as $\mathrm{Im}(n_{\mathrm{ref}}) > 0$.
    \label{fig:Xenon_nref}}
\end{figure}

The index of refraction for a single element can be related to atomic scattering factors $f_{1}(\omega)$ and $f_{2}(\omega)$ (available from the Lawrence Berkeley Lab database \cite{lbloptics}), and calculated via
\begin{equation}
    n_{\mathrm{ref}}(\omega) = 1 - \frac{r_{0}}{2 \pi} \bigg ( \frac{h c}{\omega} \bigg)^2 n_{A} (f_{1}(\omega) - i f_{2}(\omega)), \label{eq:nref}
\end{equation}
where $n_{A}$ is the number density of atoms, and $r_{0} = 2.82\times 10^{-15}$ m is the classical radius of the electron. For molecular detectors, the scattering factors for each atom are added\footnote{For helium gas, the index of refraction using Eq. \ref{eq:nref} is different than the index of refraction data given by \cite{lbloptics} for energies above $\sim$ 5 keV. In this discrepancy, we use the explicit index of refraction data from \cite{lbloptics}.}. Fig. \ref{fig:Xenon_nref} gives an example of the index of refraction over the range of energies for which we are interested. While in general the electric permittivity $\varepsilon(\omega, \vec{k})/\varepsilon_{0}=n^2_{\mathrm{ref}}(\omega, \vec{k})$ (not to be confused with the kinetic mixing parameter $\epsilon$) depends on the dark photon energy $\omega$ and three-momentum $\vec{k}$, for the dark photon energies and momentum transfer considered here the dipole approximation is a good approximation, and consequently the dependence of the electric permittivity on $\vec{k}$ is suppressed. A further discussion on this point can be found in Appendix \ref{app:electric-permittivity}.

The absorption rate for a single dark photon within our material is $\Gamma = -|\epsilon_{\mathrm{eff}}|^2 \mathrm{Im}(\Pi_{T}(\omega))/\omega$ \cite{PhysRevLett.111.041302}. Our rate of dark photon events within the detector is therefore

\begin{equation}
    R_{A'} = |\epsilon_{\mathrm{eff}}|^2 V_{\mathrm{det}} \Phi_{A'} \frac{-\mathrm{Im}(\Pi_{T}(m_{\chi}))}{m_{\chi}v_{A'}}, 
\end{equation}
where $V_{\mathrm{det}}$ is the volume of our detector and we have used the fact that $\omega = m_{\chi}$ for our flux of dark photons.  As $m_{\chi} \gg m_{A'}$, $v_{A'} \approx c$.

We can see from Eq. \ref{eq:eps_eff} that for $m^2_{A'} \gg \Pi_{T}$, we have $\epsilon_{\mathrm{eff}} = \epsilon$, while for $m^{2}_{A'} \ll \Pi_{T}$, the effective kinetic mixing is suppressed by $\mathscr{O}(m^2_{A'}/\Pi_{T})$. Avoiding this suppression leads us to look at low density detectors where $n_{\mathrm{ref}}$ is close to unity. We also note that there is an enhancement when $m_{A'}^2 = \mathrm{Re}(\Pi_{T})$.

\begin{figure*}
    \subfloat[Liquid xenon constraints/sensitivities \label{subfig-liq-xe-mA}]{%
      \includegraphics[width=0.475\textwidth]{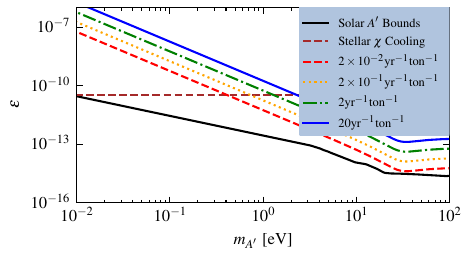}
    }
    \hfill
    \subfloat[Liquid argon constraints/sensitivities \label{subfig-ar-mA}]{%
      \includegraphics[width=0.475\textwidth]{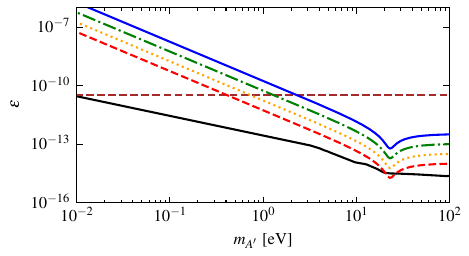}
      }

    \subfloat[Solid germanium constraints/sensitivities \label{subfig-ge-mA}]{
    \includegraphics[width = 0.475\linewidth]{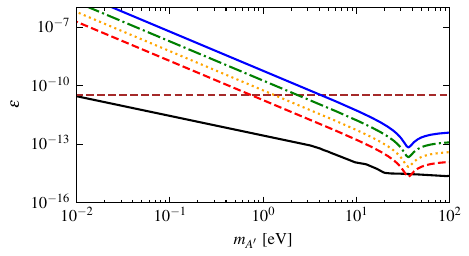}
    }
    \hfill
    \subfloat[Gaseous xenon constraints/sensitivities ($\rho_{Xe} = 5 \, \mathrm{kg \, m}^{-3}$) \label{subfig-gas-xe-mA}]{%
      \includegraphics[width=0.475\linewidth]{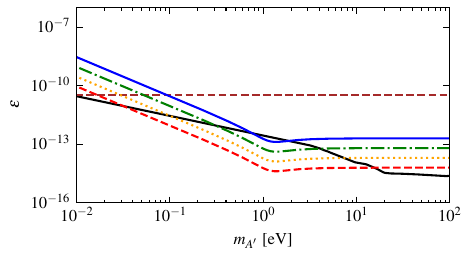}
    }

    \subfloat[Helium constraints/sensitivities ($\rho_{He} = 0.2 \, \mathrm{kg \, m}^{-3}$) \label{subfig-he-mA}]{ \includegraphics[width = 0.475\linewidth]{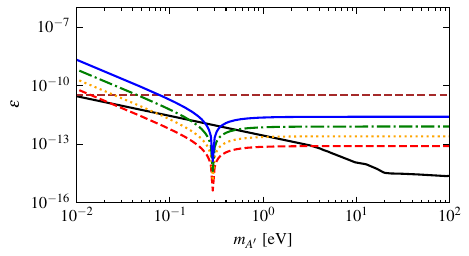}}
    \hfill
    \subfloat[Sulfur hexafluoride constraints/sensitivities ($\rho_{SF_{6}} = 6.2\, \mathrm{kg \, m}^{-3}$) \label{subfig-SF6-mA}]{%
      \includegraphics[width=0.475\textwidth]{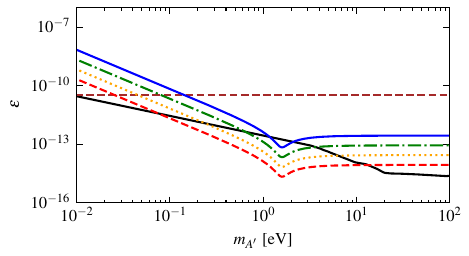}
      }
    \hfill

        \caption{Exclusion plots for various materials and assumed sensitivities for $m_{\chi} = 1$ keV, $f_i$ = 0.1, and the annihilation cross section taken to be $\langle\sigma v\rangle_{\max}$. Included are existing dark photon bounds from \cite{PhysRevD.104.095029} and $\chi$ bounds from \cite{Davidson:2000hf}. \label{money-plot}}
  \end{figure*}

\section{Results}
Our model is dependent upon 5 parameters

\begin{equation}
    (m_{A'}, \epsilon, m_{\chi}, g_{D}, f_{i}).
\end{equation}

Oftentimes, we will mention $\langle \sigma v \rangle$ instead of $g_{D}$. We can relate the two up to an order 1 factor via \cite{PhysRevD.82.083525}
\begin{equation}
    \langle\sigma v \rangle \sim \frac{\pi \alpha_{D}^2}{m_{\chi}^2},
\end{equation}
where $\alpha_{D} = g_{D}^2/4\pi$.

In Fig. \ref{money-plot}, we fix the value of $f_{i} = 0.1$ and $m_{\chi} = 1$~keV, and let $\langle \sigma v \rangle = \langle \sigma v \rangle_{\max}$ be given by Eq. \ref{eq:sigmav_max} with our chosen $m_{\chi}$ value. We vary $m_{A'}$ from 0.01 to 100 eV, and find the corresponding value of $\epsilon$ which gives the {\em a priori} event rate.


If we were to consider to consider arbitrary $\langle \sigma v \rangle$, then because $\epsilon \sim \Phi_{A'}^{-1/2}$, our sensitivity would roughly scale as

\begin{equation}
    \log(\epsilon') \sim \frac{1}{2} \bigg| \log \bigg( \frac{\langle \sigma v \rangle}{\langle \sigma v \rangle_{\max}} \bigg) \bigg|.
\end{equation}

\begin{figure*}
    \subfloat[Liquid xenon constraints/sensitivities ($m_{A'}$ = 25 eV) \label{subfig-liq-xe-mchi}]{%
      \includegraphics[width=0.475\textwidth]{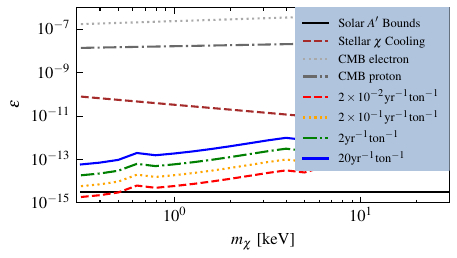}
    }
    \hfill
    \subfloat[Gaseous xenon constraints/sensitivities ($\rho_{Xe} = 5 \, \mathrm{kg \, m}^{-3}$ ; $m_{A'}$ = 1 eV) \label{subfig-gas-xe-mchi}]{%
      \includegraphics[width=0.475\linewidth]{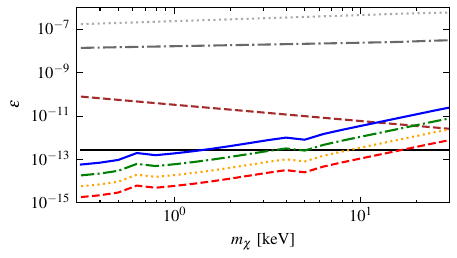}
    }
    
    \subfloat[Helium constraints/sensitivities ($\rho_{He} = 0.2 \, \mathrm{kg \, m}^{-3}$ ; $m_{A'} = 0.3$ eV) \label{subfig-he-mchi}]{ \includegraphics[width = 0.475\linewidth]{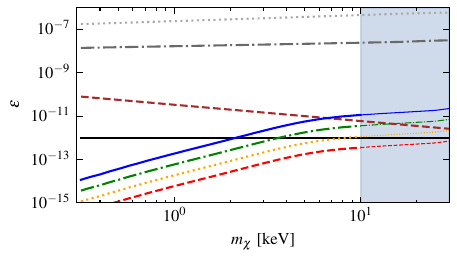}}
    \hfill
    \subfloat[Sulfur hexafluoride constraints/sensitivities ($\rho_{SF_{6}} = 6.2\, \mathrm{kg \, m}^{-3}$ ; $m_{A'}$ = 2 eV) \label{subfig-SF6-mchi}]{%
      \includegraphics[width=0.475\textwidth]{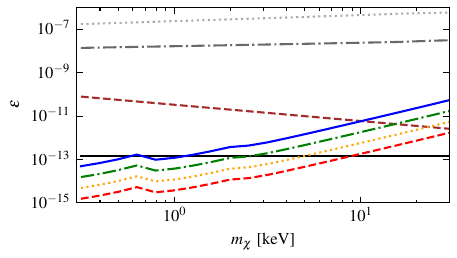}
      }    
        \caption{Exclusion plots for various materials and assumed sensitivities for $f_i$ = 0.1, with the annihilation cross section taken to be $\langle \sigma v \rangle_{\max}$. Compared to Fig.~\ref{money-plot}, here the $\chi$ mass is varied with values for $m_{A'}$ taken to be near the resonance (which is material--dependent as seen in Fig.~\ref{money-plot}). The horizontal line shows the existing dark photon bound from \cite{PhysRevD.104.095029} for our chosen $m_{A'}$. Also included are bounds from $\chi$ scattering in the early universe, obtained from \cite{buen2022cosmological}. We shade the helium results beyond 10 keV because our dipole assumption for absorption breaks down beyond this point (see Appendix \ref{app:electric-permittivity} for more details). Calculating the true sensitivity bounds in this region is beyond the scope of this paper, but using our method, the sensitivity is already subdominant to solar emitted $A^{\prime}$ by $m_{\chi} =$ 10 keV. The cooling and CMB bounds should still be valid in this shaded region.\label{money-plot-3}}
  \end{figure*}

Alternatively, in Fig. \ref{money-plot-3}, we instead fix the values of $f_{i} = 0.1$ and $m_{A'}$. We vary $m_{\chi}$ from 0.3 to 30 keV, and at every value, we set $\langle \sigma v \rangle$ to be the corresponding value of $ \langle \sigma v \rangle_{\max}$. As before, the value of $\epsilon$ which provides the desired event rate is found and graphed.





\subsection{Liquid Xenon/Argon}
The first detector materials we consider are liquid xenon and argon, which are used in low background ton-scale experiments such as dark matter direct detection \cite{XENONCollaboration:2022kmb,XENON:2018voc,DarkSide:2022dhx, DarkSide:2022knj} and neutrinoless double-beta decay (for the case of argon \cite{LEGEND:2021bnm} and xenon \cite{nEXO:2017nam}). For these noble liquids, the polarization tensor $\Pi_{T} \sim \mathscr{O}(100 \mathrm{eV}^2)$, which we can see from the kinks in $\epsilon - m_{A'}$ plots in Figs. \ref{subfig-liq-xe-mA} \& \ref{subfig-ar-mA}. Because of the relatively large (compared to gasses) in-medium effect, these liquid detectors are unable to probe any new parameter space.

\subsection{Solid Germanium}
Similar to liquid xenon and argon, germanium detectors have been used in dark matter direct detection \cite{EDELWEISS:2020fxc} and neutrinoless double-beta decay experiments \cite{Edzards:2020qdd}. 
While germanium crystals are not isotropic and so do not satisfy our assumption of isotropy, here we only consider incident dark photons with $\cal{O}($keV$)$ energies and higher. For absorption the momentum-transfer is also of this scale, and thus we expect neglecting the band structure and the anisotropy of the crystals to be a good approximation.
Again, the relatively high densities of these detectors make them unable to probe new parameter space, as can be seen in Fig. \ref{subfig-ge-mA}. 

\subsection{Gaseous Xenon}
Ton-scale gaseous xenon detectors have been proposed in the context of searching for double-beta decay such as the PANDAX-III \cite{chen2017pandax} and NEXT \cite{gomez2019status} experiments. For gases, the size of $\Pi_{T}$ depends upon the target density. We can see in Figs. \ref{subfig-gas-xe-mA} \& \ref{subfig-gas-xe-mchi} that xenon at 5 kg $\mathrm{m}^{-3}$ (the density at standard temperature and pressure) can probe new parameter space if the experiment is sensitive to rates $\mathscr{O}$(10 $\mathrm{ton}^{-1} \mathrm{yr}^{-1}$) at keV scale energies. This may be overly optimistic for double-beta decay experiments, which are designed to be most sensitive near the $Q$-value of the isotope in question (2.46 MeV for xenon-136). However, a future gasesous xenon detector with a low energy threshold and strong background discrimination could look for this signal.

\begin{figure}[hbt]
    \includegraphics[width=1\columnwidth]{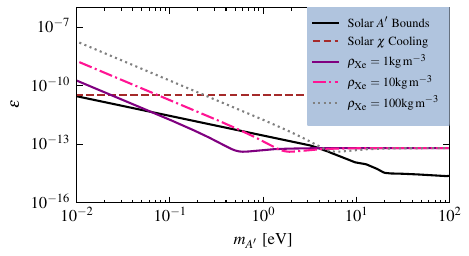}
    \caption{Exclusion and sensitivity plots for gaseous xenon detectors for a variety of gaseous xenon densities and with $m_{\chi} = 1$ keV, $f_{i}$ = 0.1, and assuming a sensitivity of 2 events $\mathrm{ton}^{-1} \mathrm{yr}^{-1}$. Included are bounds from \cite{PhysRevD.104.095029} and \cite{Davidson:2000hf}. \label{fig:Diff_Densities}}
\end{figure}

We would also like to point out Fig. \ref{fig:Diff_Densities}, in which we consider the hypothetical sensitivities of gaseous xenon detectors with the same sensitivity but different xenon densities. We see that lowering the density allows the experiment to probe lower dark photon masses.

\subsection{Helium / Sulfur Hexafluoride}
Gaseous helium and sulfur hexafluoride detectors interest us because of the proposed CYGNUS experiment \cite{Vahsen:2020pzb}, a large volume dark matter detector which contains a mixture of the two gasses. This experiment has been considered to reach sizes of 1000 $\mathrm{m}^3$ and beyond, with very low energy thresholds, making it an interesting detector to search for these boosted dark photons. We would like to make a special note of helium gas in Fig. \ref{subfig-he-mA}. Although helium has a relatively small cross-section for x-rays, that same property leads to a strong resonance in Eq. \ref{eq:eps_eff}. At a single density, the helium detector would be sensitive to a small range of dark photon masses. However, one could imagine a varying density (either in time or across modules) that would allow for an improved search over a larger range of masses.

\section{Existing constraints on $A'$ and $\chi$}


\subsection{Dark Photon Limits}
One can consider an extension to the Standard Model given only by Eq. \ref{eq:DP_Lagrangian}. This model has a rich phenomenology and many constraints on $m_{A'}$ and $\epsilon$ as shown in Figure 1 of \cite{PhysRevD.104.095029} and regularly updated in corresponding website. Note that we disregard the bounds which arise when the dark photon is the dark matter candidate, as that is different than the scenario considered here. In our region of interest ($m_{A'} = \mathscr{O}(10^{-2}-10^{1} \mathrm{eV}) )$ the most stringent constraints consider the Sun as a source of dark photons, and constrain either solar cooling \cite{an2013new} or direct detection of these emitted $A'$ \cite{PhysRevD.102.115022}.

\subsection{SM - $\chi$ Scattering}
Our model allows for $\chi$ to scatter with electrically charged SM particles by exchanging a virtual dark photon which mixes with the SM photon. We consider $m_{\chi} \sim \mathscr{O}$(1-100 keV) with virialized velocity, leading to kinetic energies of $\mathscr{O}(10^{-3} - 10^{-1}$eV). These energies are too small to probe with modern-day direct detection experiments. However, in the early universe, scattering between $\chi$ and SM particles can leave an observable imprint on the Cosmic Microwave Background (CMB)\cite{buen2022cosmological,nguyen2021observational}. We can translate constraints on cross section into constraints on $\epsilon$, $g_{D}$, $m_{\chi}$, and $f_{i}$ by assuming that $\chi$ is the only component of dark matter to interact with SM particles. One could also consider the energy injection into the early universe from the annihilation of $\chi$ as in \cite{CMBAnnihilation}, but the resulting bounds are much weaker than those from scattering.

We do not consider constraints in which $\chi$ interacts with long-range electric or magnetic fields. Although there is mixing between the SM and dark photons, the dark photon masses considered here are large enough that the effects of long-range SM fields are exponentially suppressed.

\subsection{$\chi$ Driven Stellar Cooling}
Similar to dark photons, $\chi \Bar{\chi}$ pairs can be produced in stars with temperature $T \gtrsim m_{\chi}$. The strongest of these constraints come from horizontal branch stars, which limit $q_{D} = g_{D} \epsilon/e < 2 \times 10^{-14}$ \cite{Davidson:2000hf}.

\subsection{$\chi$ - $\chi$ Scattering}
Gravitational measurements of galactic mergers and the ellipticity of galactic dark matter halos place constraints on the self-interactions of dark matter \cite{lasenby2020long}, which in our model can be translated into constraints on $m_{\chi}$, $g_{D}$, $f_{i}$, and $m_{A'}$. The simplest way to avoid these constraints is to take $f_{i} \sim \mathscr{O}(0.1)$, so that any effects of $\chi$ will be smaller than the uncertainty on the measurements.

\section{Decaying Dark Matter}
The analysis considered in this paper could similarly be used to constrain a dark matter candidate $\phi$ which would decay into dark photons ($\phi \rightarrow A' A'$). If we assume that $\phi$ is the entirety of dark matter, this model can be characterized by four parameters

\begin{equation}
    (m_{A'}, \epsilon, m_{\phi}, \Gamma_{\phi}),
\end{equation}
where $\Gamma_{\phi}$ is the decay rate of $\phi$. This will produce a galactic flux of dark photons which is given by

\begin{equation}
    \Phi_{A', \mathrm{dec}} = \frac{1}{4 \pi} \frac{2 \Gamma_{\phi}}{m_{\phi}} \int d\Omega \int_{\mathrm{LOS}} dx \rho_{\phi}(x,\theta).
\end{equation}

\begin{figure}[hbt]
    \includegraphics[width=1\columnwidth]{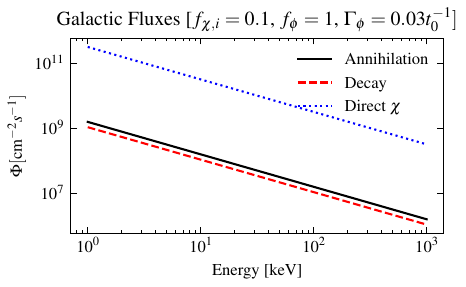}
    \caption{Comparing the galactic dark photon flux from annihilating dark matter with annihilation cross section given by Eq. \ref{eq:sigmav_max} and initial fraction $f_{i}$ = 0.1, and decaying dark matter with $\Gamma_{\phi}$ = 0.03 $t_{0}^{-1}$. For annihilation and decay, ``Energy" refers to the dark photon energy, while for the direct $\chi$ flux, it refers to $m_{\chi}$. \label{fig:Ann_Decay_Direct}}
\end{figure}

Taking a value for $\Gamma_{\phi} = 0.03 t_{0}^{-1}$, which is consistent with \cite{chen2021constraints}, we obtain a flux of dark photons comparable with the maximum allowed flux from $\chi \bar{\chi}$ annihilations. (In general, the dark photon flux could also have a contribution from extragalactic $\phi$, but for now, only the galactic contributions are compared). In Fig. \ref{fig:Ann_Decay_Direct} we compare the two fluxes, along with the direct flux of dark matter at Earth, calculated via $\Phi_{\chi} = \frac{\rho_{\chi}}{m_{\chi}} v_{\odot}$ where $v_{\odot}$ is the velocity of the Sun around the Milky Way.

\section{Conclusions}
In this paper, we have shown that for $\chi$ being a sub-component of dark matter with large annihilation cross sections, there exists currently unconstrained parameter space that can lead to visible signals in future gaseous detectors. We do not provide exclusions or projections, as that would require essential features of future experiments, such as energy thresholds and estimates of backgrounds, that are beyond the scope of this work. However, we do provide code to determine the rate of dark photon events at a particular experiment (see the GitHub link in the abstract). We also wish to point out some detector properties that would be most useful in looking for this signal

\begin{itemize}
    \item \textit{\textbf{Low Densities and Large Volumes:}} Clearly, a larger target mass allows for more dark photons to be absorbed, and we see in Fig. \ref{fig:Diff_Densities} that lower densities can be sensitive to lower $m_{A'}$ since the dark photons are predominantly transversely polarized. A variable density (with significant exposure) could scan parameter space for an even larger range of dark photon masses.

    \item \textit{\textbf{Low Energy thresholds and Good Energy Resolution:}} The flux of $A'$ is largest for small $m_{\chi}$, but this flux would only be visible if the energy threshold is below $m_{\chi}$. Moreover, the flux of dark photons would be mono-energetic. Good energy resolution and reconstruction would have these dark photon events appear as a peak, reducing the number of events necessary to be significant.

    \item \textit{\textbf{Good Spatial Resolution:}} Typical x-ray backgrounds, which may come from outside the detector or detector components, will mainly occur on the detector edges and be unable to penetrate to the center of the fiducial volume. On the other hand, the small interaction probability of dark photons would make them equally likely to appear anywhere within the detector, so topological discrimination could be used to reduce the background.
\end{itemize}


\section*{Acknowledgements}

The authors thank Kaustubh Agashe, Haipeng An, Steve Elliott, Yoni Kahn, Doojin Kim, Ralph Massarczyk, and Sam Watkins for comments and discussions.
The work of MG is supported by the LDRD program at Los Alamos National Laboratory and by the U.S. Department of Energy, Office of High Energy Physics, under 
Contract No. DE-AC52-06NA25396. RAG, VM, and IMS are supported by the U.S. Department of Energy Office of Science, Office of High Energy Physics, under Award Number DE-SC0020262. KH was supported in part by the VT Center for Neutrino Physics REU program: NSF award number 2149165. 

\appendix

\section{Time Evolution of $\chi$ and Assumptions} \label{appendix: DM Time Evolution}
We consider the parametrization of the $\chi$ density given by Eq. \ref{eq:rho_chi_decomp} (here called the "Diffusion Limit"). Using Eq. \ref{eq:AnnTimeEvo}, we need to integrate over the galaxy to find the time evolution of the scaling factor.

\begin{equation}
    \frac{d f_{\chi}}{d t} \int \rho_{\mathrm{NFW}}(\textbf{x}) dV = \frac{-\langle\sigma v\rangle f_{\chi}^2(t)}{m_{\chi}} \int \rho^2_{\mathrm{NFW}}(\textbf{x}) dV
\end{equation}

This differential equation is solved by Eq. \ref{eq:chi_frac}.

An alternative approach to the time evolution of $\chi$ is what we call Spatially-Independent Evolution (SIE). In this case, we consider the initial distribution to be proportional to the NFW distribution $\rho_{\chi,\mathrm{SIE}}(r, t=0) = f_{i,\mathrm{SIE}}\rho_{\mathrm{NFW}}(r)$, and then each location in space evolves independent of other locations in space, according to Eq. \ref{eq:AnnTimeEvo}. This is solved by

\begin{equation}
    \rho_{\chi,\mathrm{SIE}}(r,t) = \frac{f_{i, \mathrm{SIE}} \times \rho_{\mathrm{NFW}}(r) m_{\chi}}{m_{\chi} + f_{i,\mathrm{SIE}} \times \rho_{\mathrm{NFW}}(r) \langle \sigma v \rangle t}.
\end{equation}

\begin{figure}[hbt]
    \includegraphics[width=1\columnwidth]{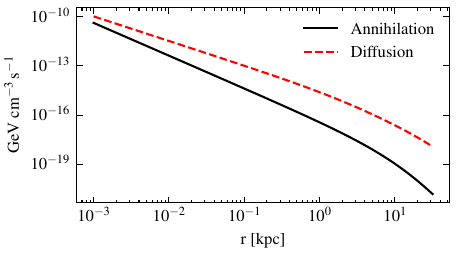}
    \caption{Comparing the two equations in Eq. \ref{eq:drho_dt_compare} as a function of radius r from the galactic center, with $f_{i} = 0.1$ and $m_{\chi} = 1$ keV.
    \label{fig:Annihilation_v_Diffusion}}
\end{figure}

Intuitively, SIE is the correct approach when annihilation happens very quickly, whereas the Diffusion Limit is motivated when $\chi$ diffuses fast enough to correct for any changes due to annihilations. To give a quantification of this comparison, we will compare

\begin{equation}
    v_{\mathrm{NFW}}(r) f_{i}| \nabla \rho_{\mathrm{NFW}}(r) | \, \, \mathrm{vs} \, \, \frac{\langle\sigma v \rangle f^2_{i} \rho^2_{\mathrm{NFW}}(r)}{m_{\chi}}, \label{eq:drho_dt_compare}
\end{equation}
where the left expression is the characteristic change in density due to diffusion, and the right is due to annihilations. We define this characteristic velocity as

\begin{equation}
    v_{\mathrm{NFW}}(r) = \sqrt{\frac{G M_{\mathrm{enc}}(r)}{r}},
\end{equation}
where $M_{\mathrm{enc}}$ is enclosed mass in a sphere centered at the galactic center of radius $r$. Along with the dark matter density, we also need to consider the baryonic density \cite{Berlin:2023qco,1990ApJ...356..359H}

\begin{equation}
    \rho_{B}(r) = \frac{\rho_{B0} r^4_{0}}{r (r+r_0)^3},
\end{equation}
where $\rho_{B0}=$ 26 GeV $\mathrm{cm}^{-3}$ and $r_{0} = 2.7$ kpc. We now have the information needed to compare the two expressions in Eq. \ref{eq:drho_dt_compare}, and we will consider $\langle \sigma v \rangle = \langle \sigma v \rangle_{\max}$ as defined in Eq. \ref{eq:sigmav_max}. As can be seen in Fig. \ref{fig:Annihilation_v_Diffusion}, the diffusion occurs faster than annihilation. For this reason, we assume the Diffusion Limit is a good approximation, although a more detailed simulation would be needed to capture all of the physics.

\section{Non-relativistic Dark $\chi$ Annihilation and Dark Photon Polarizations}
This Appendix considers non-relativistic $\chi\overline{\chi}$ annihilation to two dark photons. In the limit $E_\chi \simeq m_\chi \gg m_{A'}$, the final state is found to be dominated by transversely polarized dark photons. Production of longitudinally polarized dark photons is highly suppressed. 

\label{app:dark-photon-polarizations}
In order to demonstrate the dominance of transverse polarizations, we subtract off the transversely polarized contribution from the total rate and compare the two.
The process of $\chi(k_{1}) \bar{\chi(k_{2})} \rightarrow A^{\prime,\nu}(p1)A^{\prime,\mu}(p_2)$ at tree-level consists of the matrix element
 \begin{equation}
    \begin{split}
     M = i g_{D}^2 \bar{v(k_{2})} \bigg( \gamma^{\mu} \epsilon^{*}_{\mu}(p_{2}) \frac{\slashed{p_{2}} - \slashed{k_{2}} + m_{\chi}}{t - m_{\chi}^2}  \gamma^{\nu} \epsilon^{*}_{\nu}(p_{1}) \\
     + \gamma^{\nu} \epsilon^{*}_{\nu}(p_{1}) \frac{\slashed{p_{1}} - \slashed{k_{2}} + m_{\chi}}{u - m_{\chi}^2}  \gamma^{\mu} \epsilon^{*}_{\mu}(p_{2}) \bigg) u(k_{1}),
     \end{split}
 \end{equation}
where $t$ and $u$ are the Mandelstam variables and $\epsilon_{\mu}(p)$ is a polarization vector, not to be confused with the kinetic mixing parameter. To compute the cross section of this process, we find $|M|^2$ and sum over the outgoing polarizations. 
We know that for a massive vector boson, the sum over polarizations (denoted by $\lambda$) returns
\begin{equation}
    \sum_{\lambda} \epsilon_{\lambda}^{*\mu}(p) \epsilon_{\lambda}^{\nu}(p) = \frac{p^{\mu} p^{\nu}}{p^2} - g^{\mu \nu}.
\end{equation}
We next define the longitudinal polarization to be
\begin{equation}
    \epsilon_{L}(p) = \frac{1}{\sqrt{\omega^2 - |\overrightarrow{p}|^2}} \bigg( |\overrightarrow{p}|, \omega \frac{\overrightarrow{p}}{|\overrightarrow{p}|} \bigg).
\end{equation}

It can be shown that
\begin{equation}
    \begin{split}
    \epsilon_{L}^{\mu}(p) \epsilon_{L}^{\nu}(p) = \delta^{\mu 0} \delta^{\nu 0} \bigg( \frac{p^2}{\omega^2 - p^2} \bigg) - \delta^{\mu 0} \bigg(\frac{\omega p^{\nu}}{p^2} \bigg) \\
    - \delta^{\nu 0} \bigg(\frac{\omega p^{\mu}}{p^2} \bigg) + \frac{p^{\mu} p^{\nu}}{\omega^2-p^2} + \frac{p^{\mu}p^{\nu}}{p^2}.
    \end{split}
\end{equation}

When we apply the on-shell condition, the first term will be $\mathscr{O}(m_{A^{\prime}}^{2}/m_{\chi}^2)$, which is small enough that we will ignore it. To find final value of the polarization sum, it is useful to define a new four-vector $\eta^{\mu}(\omega, p)$

\begin{equation}
    \eta^{\mu}(\omega,p) = \omega (1,0,0,0) - p^{\mu}.
\end{equation}

Thus we can define

\begin{equation}
    \begin{split}
    \sum_{\mathrm{transverse}} \epsilon_{\lambda}^{*\mu}(p) \epsilon_{\lambda}^{\nu}(p) =  - g^{\mu \nu} -\frac{\eta^2 p^{\mu} p^{\nu}}{(\eta \cdot p)^2} \\
    + \frac{\eta^{\nu} p^{\mu}}{(\eta \cdot p)}+ \frac{\eta^{\mu} p^{\nu}}{(\eta \cdot p)} + \mathscr{O}(m_{A^{\prime}}^{2}/m_{\chi}^2),
    \end{split}
\end{equation}
(this is easily implemented using FeynCalc \cite{Mertig:1990an,Shtabovenko:2016sxi,Shtabovenko:2020gxv}). If we work in the low-velocity limit, expanding in both $m_{A^{\prime}}$ and $K$ defined as $s = 4 (m_{\chi} + K)^2$, we find,

\begin{equation}
    \frac{1}{4}\sum_{\mathrm{spin}} (\sum_{T+L} |M|^2 - \sum_{T} |M|^{2}) = \mathscr{O}\bigg(\frac{K^{2}}{m^2_{\chi}}, \frac{m_{A^{\prime}}^2}{m^2_{\chi}} \bigg),
\end{equation}
where the term outside parentheses is to average over the incoming fermion spins, and the latter sums are the polarization sums with and without the longitudinal modes. Also, up to this level of precision

\begin{equation}
    \frac{1}{4} \sum_{\mathrm{spin}} \sum_{T} |M|^2 = 4g_{D}^4 + 16 g_{D}^4 \frac{K}{m_{\chi}} + \mathscr{O}\bigg(\frac{K^{2}}{m^2_{\chi}}, \frac{m_{A^{\prime}}^2}{m^2_{\chi}} \bigg).
\end{equation}

\section{Electric Permittivity}
\label{app:electric-permittivity}

In general the electric permittivity $\varepsilon$ is a function of both the incident (dark) photon 
energy $\omega$ as well as its three momentum $\vec{k}$. For most of the energies considered in this work, however, the dipole approximation $e^{i \vec{k} \cdot \vec{x}} \simeq 1$ is valid and the dependence of the electric permittivity on $\vec{k}$ can be neglected to leading order \cite{Bethe:1957ncq}. The reasons are as follows. 

Bound-state to bound-state transitions are necessarily for low incident energies, 
since axiomatically,  $\omega \simeq |\vec{k}|$ is bounded by the atom's ionization energy, and since very little energy is transferred to the kinetic motion of the recoiling atom (because atoms are heavy). 
Thus $\omega \simeq |\vec{k}| \simeq E_{nn'}  \sim (Z \alpha_{em})^2 m_e$, where 
$E_{nn'}$ is the energy of the transition. With 
$x \simeq a_{nn'}$, i.e. the Bohr radius of the electron states in question, and as 
$a_{nn'} \simeq 1/(Z \alpha m_e)$, one sees that $k x \ll 1$ provided $Z \alpha \ll 1$ \cite{Bethe:1957ncq}. Moreover, bound-state to bound-state electrons transitions are predominately by outer electrons, for which the effective electric charge $Z \approx 1$. (This argument is also presented in \cite{An:2013yua}).

For ionization, Bethe and Salpeter demonstrate that these processes are dominated by inner electrons (K and L shells), and this is the key point (Secs. 69-71,\cite{Bethe:1957ncq}). Because these electrons see a higher $Z$ charge, they therefore have a smaller Bohr radius and consequently $kx$ remains small. Namely, if $\omega \ll Z \alpha m_e$ (that is, the dark photon wavelength is larger than the radius of that inner electron) then retardation effects can be neglected and consequently $e^ {i k x} =1$ to a good approximation (Section 69, Eq. 69.9, \cite{Bethe:1957ncq}).\\

\vspace{-.17cm}

For argon, $Z \alpha m_e \approx 65$ keV, for fluorine $Z \alpha m_e \approx 32$ keV, for germaniun $Z \alpha m_e \approx 116$ keV, 
for sulphur $Z \alpha m_e \approx 60$ keV, and  
for xenon $Z \alpha m_e \approx 200$ keV. For these elements, our analysis is 
accordingly restricted to $\omega \lsim 30$ keV to maintain the dipole approximation. 
For helium $Z \alpha m_e \approx 7$ keV, however, such that the dipole approximation becomes invalid at energies $\gsim$ 10 keV. In Fig.~\ref{money-plot-3}c, the region above $10$ keV is highlighted to indicate the breakdown of the dipole approximation there.

\section{Comparison to Solar Bounds}
The most stringent bounds in our region of interest come from the solar production of dark photons \cite{an2013new,PhysRevD.102.115022}. We think it is important to discuss the fundamental differences between solar production and boosted production via $\chi$ annihilation at the Galactic Center.

First, in solar production, the flux of dark photons necessarily goes to zero as the dark photon mass becomes small. This is because SM currents are the source of the $A'$, and the large solar in-medium effects suppresses the effective mixing as seen in Eq. \ref{eq:eps_eff}. Alternatively, $\chi$ annihilation is scaled by $m_{\chi}$ and $g_{D}$, not $\epsilon$ or $m_{A'}$ (assuming $m_{A'} \ll m_{\chi}$), so the flux of dark photons remains to arbitrarily small dark photon masses.

Second, for solar production, dark photons are primarily in the longitudinal polarization, which have an absorption rate linearly dependent on the density \cite{PhysRevLett.111.041302}. This means that gaseous detectors would not offer an improvement over liquid detectors for equal target masses. This is quite different than the transversely dominated flux from $\chi$ annihilations, for which we see a significant improvement at gaseous detectors.

\bibliographystyle{apsrev4-1.bst}
\bibliography{bibliography.bib}

\end{document}